\documentclass[a4paper,12pt]{article}
\usepackage{amsmath,amsfonts,amssymb}
\usepackage{graphicx}

\begin{document}

\begin{center}
{\LARGE\bf Soft wall model for a holographic superconductor}
\end{center}

\begin{center}
{\large S. S. Afonin and I. V. Pusenkov}
\end{center}

\begin{center}
{\small V. A. Fock Department of Theoretical Physics,
Saint-Petersburg
State University, 1 ul. Ulyanovskaya, St. Petersburg, 198504, Russia\\
Email: \texttt{afonin@hep.phys.spbu.ru, i.pusenkov@spbu.ru} }
\end{center}

\begin{abstract}
We apply the soft wall holographic model from hadron physics to a
description of the high-$T_c$ superconductivity. In comparison
with the existing bottom-up holographic superconductors, the
proposed approach is more phenomenological. On the other hand, it
is much simpler and has more freedom for fitting the conductivity
properties of the real high-$T_c$ materials. We demonstrate some
examples of emerging models and discuss a possible origin of the
approach.
\end{abstract}

\section{Introduction}

Recently the holographic superconductors have attracted a lot of
attention (see, e.g., reviews~\cite{horowitz,cai}). They represent
certain AdS/CFT models containing black holes and describe some
important features of superconductivity. The active research in
this direction was inspired by an observation made in
Ref.~\cite{HHH} that such models could describe the high-$T_c$
superconductivity discovered in some cuprates and other materials.
This spectacular phenomenon in condensed matter has still not been
understood~\cite{carlson}. It is not described by the standard BCS
theory (a theory of weak inter-fermion interactions mediated by
phonons) since it is caused by a strong coupling between
fermions~\cite{legett}. Due to the strong coupling the
establishing of the ground state of the system becomes a hard
problem. This state is not the Fermi liquid as in the BCS theory.
One of possibilities for modelling the ground state in the
high-$T_c$ cuprates is the two-dimensional "tomographic Luttinger
liquid"~\cite{anderson}. It is interesting that the expected
ground state of the finite-density QCD (characterizing by the
color superconductivity) might have a similar
nature~\cite{shuster}.

The practical applications of the gauge/gravity duality are not
restricted by the strongly coupled gauge theories. The conformal
behavior near the critical points of phase transitions suggested
the idea to apply the methods of AdS/CFT correspondence to
strongly coupled condensed matter systems~\cite{hartnoll}. In
particular, this gave tantalizing prospects of finding a
description for the high-$T_c$ superconductivity in terms of a
dual gravitational theory.

The simplest qualitative theory for holographic superconductor was
constructed in Ref.~\cite{HHH}. It represents an Abelian--Higgs
bottom-up holographic model with black hole. The model shares some
similarities with the Ginzburg--Landau theory of
superconductivity. All other bottom-up holographic superconductors
proposed in the literature can be viewed as extensions of the
model of Ref.~\cite{HHH}. The existing holographic models, both in
the bottom-up and in the top-down approaches, give only a rough
qualitative description of the real high-$T_c$ materials. From the
point of view of a condensed matter phenomenologist, it would be
interesting to have a simple holographic superconductor with
several adjustable parameters which would allow to fit the
physical characteristics of a concrete superconducting material.
In the present Letter, we propose a design for such a model.

The idea of our approach is borrowed from the soft wall (SW)
holographic model in hadron physics~\cite{son2}. The purpose of
construction of the SW model was essentially the same ---
advancing in quantitative holographic description of real
experimental data. The SW model is well tunable for description of
the hadron spectra and related phenomenology, albeit it does not
describe (at least in its simplest version) the spontaneous chiral
symmetry breaking in QCD. Our SW holographic superconductors will
be also quite flexible for fitting the observable properties of
high-$T_c$ superconductors, first of all the behavior of optical
conductivity. However, we need to pay a similar price --- the
simplest SW superconductor does not describe the superconducting
phase transition. Embedding this description seems to require a
certain complication. But if we are interested in a description of
experimentally measurable quantities, the approach that we propose
should be considered as a step forward to building a
phenomenologically useful holographic description of high-$T_c$
superconductors. A pleasant feature of our approach is that the
emerging models are much simpler than the original model of
Ref.~\cite{HHH} which before has been regarded (in the probe
limit) as the simplest holographic superconductor.

The structure of the paper is as follows. In order to make the
text self-contained and to help in comparing our approach with the
standard one, we remind the reader the idea of the first bottom-up
holographic superconductor~\cite{HHH} in Section~2. Our SW
superconductor is introduced in Section~3. Section~4 contains
discussions and we conclude in Section~5. All plots are presented
in Appendix.

\section{The simplest standard model of a holographic superconductor}

We will recall very briefly the basics of the first holographic
model for the high-$T_c$ superconductivity. This model was
constructed as a 3+1 dimensional Einstein gravitational theory
with a negative cosmological constant. By assumption, it is dual
to a 2+1 dimensional superconductor. The dimensionality was
dictated by the experimental fact that the high-$T_c$ cuprates and
other high-$T_c$ materials are usually layered and much of the
physics is effectively 2+1 dimensional. The model includes a
$U(1)$ gauge field interacting (following an analogy with the
Landau--Ginzburg theory of superconductivity) with a complex
scalar field $\psi$. The action of the model is~\cite{HHH}
\begin{equation}
\label{1}
S=\int d^4x\sqrt{-g}\left(R+\frac{6}{L^2}-\frac14F_{\mu\nu}F^{\mu\nu}-|\nabla\psi-iqA\psi|^2-m^2\psi^2\right),
\end{equation}
where $R$ is the scalar curvature, $F_{\mu\nu}=\nabla_\mu A_\nu -
\nabla_\nu A_\mu$, L is the AdS radius. The most of the
interesting physics appears already in the probe limit:
$q\rightarrow\infty$ with $qA_\mu$ and $q\psi$ fixed. This limit
simplifies considerably the model because the backreaction of the
fields on the metric is neglected. As a background metric one
considers the planar Schwarzschield AdS black hole
\begin{equation}
\label{2}
ds^2=-f(r)dt^2+\frac{dr^2}{f(r)}+r^2(dx^2+dy^2),
\end{equation}
\begin{equation}
\label{3}
f(r)=\frac{r^2}{L^2}\left(1-\frac{r_0^3}{r^3}\right).
\end{equation}
The Schwarzschield radius $r_0$ yields the Hawking temperature,
\begin{equation}
\label{4}
T=\frac{3r_0}{4\pi L^2}.
\end{equation}

The black hole in this model is unstable and forms "scalar hair".
To see this one looks for static translationally invariant
solutions: $A_r=A_x=A_y=0$, $A_t=\phi(r)$, $\psi=\psi(r)$. The
equations of motion for $\phi$ and $\psi$ with certain boundary
conditions give the relevant solutions. In Ref.~\cite{HHH}, the
case $m^2=-2/L^2$ was considered (we will set $L=1$ in what
follows). In this case, the solutions regular at the horizon have
the following asymptotics at infinity,
\begin{equation}
\label{5}
\psi=\frac{\psi^{(1)}}{r}+\frac{\psi^{(2)}}{r^2}+\dots,\qquad \phi=\mu-\frac{\rho}{r}+\dots,
\end{equation}
where $\mu$ and $\rho$ are interpreted as the chemical potential
and charge density of the dual field theory. If we set
$\psi^{(1)}=0$, the behavior of the condensate $\langle O_2
\rangle=\psi^{(2)}$ as a function of temperature is depicted in
Fig.~1 (see Appendix). As $T\rightarrow T_c$, the behavior is
$\langle O_2 \rangle\sim(1-T/T_c)^{1/2}$ as in the
Landau--Ginzburg theory~\cite{HHH}.

The plot in Fig.~1 is obtained in the following way. By symmetry,
$\psi^{(2)}$ is a function of $r_0/\sqrt{\rho}$. One solves
numerically the system of equations of motion at fixed $\rho$ and,
using~\eqref{4}, the critical temperature $T_c$ is defined as
$\psi^{(2)}(T_c/\sqrt{\rho})=0$. Thus, $T_c\sim\sqrt{\rho}$ and
one can draw the plot in Fig.~1 in units of $T_c$.

An important measurable quantity in superconductors is the optical
conductivity (the conductivity as a function of frequency). By
symmetry, it is enough to analyze the conductivity in the $x$
direction. Consider the perturbations of the vector field in the
form
\begin{equation}
\label{6}
A_\mu=A_x(r)e^{i\omega t}.
\end{equation}
The linearized equation of motion reads
\begin{equation}
\label{7}
A_x''+\frac{f'}{f}A_x'+\left(\frac{\omega^2}{f^2}-\frac{2\psi^2}{f}\right)A_x=0.
\end{equation}
The causal behavior is provided by the ingoing wave boundary
condition at the horizon~\cite{HHH},
\begin{equation}
\label{7b}
A_x\sim f^{-i\omega/3r_0}.
\end{equation}
The asymptotic behavior at infinity is
\begin{equation}
\label{8}
A_x=A_x^{(0)}+\frac{A_x^{(1)}}{r}+\dots.
\end{equation}
Since the behavior~\eqref{8} holds near the AdS boundary the
gauge/gravity correspondence~\cite{witgub} tells us that
$A_x^{(0)}$ must be identified with the source and $A_x^{(1)}$ is
dual to the induced current $J_x$. Then the definition of the
conductivity and the ansatz~\eqref{6} result in
\begin{equation}
\label{9}
\sigma(\omega)=\frac{J_x}{E_x}=\frac{J_x}{-\dot{A}_x^{(0)}}=\frac{-iA_x^{(1)}}{\omega A_x^{(0)}}.
\end{equation}
The typical behavior of real and imaginary parts of
$\sigma(\omega)$ at $m^2=-2$ is shown in Figs.~2 and~3. This
behavior looks similar to the experimental data on the AC
conductivity for the high-$T_c$ superconductors
reproduced\footnote{The gate voltage in those figures plays the
role of the current density which fixes the temperature scale in
the model.} in Figs.~4 and~5. The optical conductivity is constant
above $T_c$, below $T_c$ it develops a gap at some frequency
$\omega_g$ which can be identified with the minimum of
$\text{Im}[\sigma(\omega)]$~\cite{roberts}. The gap is the most
pronounced in the limit $T\rightarrow 0$. In this limit
$\sqrt{\langle O_2\rangle}/T_c\simeq7$ (see Fig.~1). In the same
limit one can calculate $\omega_g/\sqrt{\langle
O_2\rangle}\simeq1.2$. Both relations turn out to be weakly
dependent on the choice of parameters. Excluding $\sqrt{\langle
O_2\rangle}$, one arrives at an approximately (within 10\%)
universal relation
\begin{equation}
\label{10}
\frac{\omega_g}{T_c}\simeq8.4.
\end{equation}
The obtained ration~\eqref{10} is quite remarkable. First, it is
close to that in the high-$T_c$ superconductors~\cite{gomes}.
Second, the comparison with the BCS value $\omega_g\simeq3.5T_c$
shows that the holographic model under consideration seems to
describe a system at strong coupling.

At this point we should make a remark that will be crucial for
constructing our model in the next Section. The black
hole~\eqref{2} is invariant under the rescaling
\begin{equation}
\label{11}
r\rightarrow\lambda r, \qquad (t,x,y)\rightarrow (t,x,y)/\lambda, \qquad r_0\rightarrow\lambda r_0.
\end{equation}
This entails the invariance $\omega\rightarrow\lambda\omega$ and
(by~\eqref{4}) $T\rightarrow\lambda T$ leading to the scale
invariance of the ratio $\omega/T$. The positions of minima
$\left(\frac{\omega}{\lambda T_c}\right)_{\text{min}}$ in Fig.~3
can be viewed as $\frac{1}{\lambda}\frac{\omega_g}{T_c}$. Thus,
the ratio~\eqref{10} can be obtained from any plot in Fig.~3 just
multiplying by $\lambda$ the position of the corresponding
minimum.

The observation above hints at the possibility to construct the
holographic superconductors without condensates. Such models
should respect the scaling symmetry~\eqref{11} and reproduce the
experimental plots in Figs.~4 and~5 as close as possible. After
that one can simply extract the ratio~\eqref{10}. Below we build
some examples for such models.

\section{Soft wall holographic superconductor}

There exist common features in constructing the bottom-up
holographic models for superconductivity and for hadron physics.
The solution of equations of motion should yield the optical
conductivity in the former case and the correlation functions
(with ensuing hadron spectrum) in the latter one. The first
bottom-up model for hadrons was proposed in Ref.~\cite{son1}. The
action of this model looks like a five-dimensional extension
of~\eqref{1}. The condensation of the scalar field $\psi$
described the spontaneous chiral symmetry breaking in QCD. An
axial-vector field was also introduced in order to get the chiral
dynamics and the mass splitting between the vector and axial
mesons. Thus one deals with very similar equations but imposes
completely different requirements on the solutions. The important
differences in the bottom-up description of hadrons in
Ref.~\cite{son1} are as follows: a) In the probe limit, the
background metric is pure AdS (unless one is interesting in some
finite-temperature effects); b) The infrared cutoff
$r_{\text{IR}}$ is introduced by hands to provide the mass scale
(for this reason the model of Ref.~\cite{son1} is referred to as
the "hard wall" model); c) The scalar field $\psi$ condenses
first, i.e. the vector fields do not enter the equation of motion
for $\psi$.

The hard wall model describes well the chiral dynamics but gives
very rough predictions for the hadron spectrum. In order to
improve the second aspect the so-called "soft wall" holographic
model was introduced in Ref.~\cite{son2}. The action of the
simplest SW model describing the vector spectrum is
\begin{equation}
\label{12}
S_{\text{SW}}=\int d^5x\sqrt{|g|}\,e^{-2\varphi(ar)}\left(-\frac14F_{MN}F^{MN}\right).
\end{equation}
Now there is no infrared cutoff and the mass scale is provided by
the parameter $a$. The experimental (or theoretically expected)
spectrum can be fine-tuned by a choice of the dilaton background
$\varphi$. Simultaneously, one may reproduce much better the
structure of the Operator Product Expansion of the QCD correlation
functions. In principle, the dilaton background may be formally
(i.e. neglecting a surface term) eliminated by the substitution
$A_{M}\rightarrow e^{\varphi} \tilde{A}_M$. The price to pay is
the appearance of $r$-dependent mass term for $\tilde{A}_M$.
However, the model can be reformulated in a gauge-invariant way if
we assume that this term emerges from condensation of some scalar
field $\psi$~\cite{nowall}. In particular, the standard choice
$\varphi\sim r^{-2}$~\cite{son2} leads to the (five-dimensional
extension of) field part of action~\eqref{1} with
$m^2=-4$~\cite{nowall}. In such a "no-wall" model it is assumed
that the field $\psi$ condenses first and only after that one
analyses the ensuing phenomenology.

We are going to use the analogy from the hadron physics as the
starting point for construction of the SW holographic
superconductor. Consider the action
\begin{equation}
\label{13}
S=\int d^4x\sqrt{-g}\,e^{-2h\left(\frac{r}{r_0}\right)}\left(-\frac14F_{\mu\nu}F^{\mu\nu}\right).
\end{equation}
The background metric is the black hole one~\eqref{2}. By
assumption, the dilaton profile $-2h\left(\frac{r}{r_0}\right)$ emerges due to
condensation of some scalar field and satisfies the scaling
invariance~\eqref{11}. Making use of the ansatz~\eqref{6} in the
equation of motion, we arrive at the equation
\begin{equation}
\label{14}
A_x''+\left(\frac{f'}{f}-2h'\right)A_x'+\frac{\omega^2}{f^2}A_x=0.
\end{equation}

As in the "no-wall" model for hadrons, one can eliminate the
dilaton background in~\eqref{13} by the substitution
$A_x\rightarrow e^h A_x$ and come to the action
\begin{equation}
\label{15}
S=\int d^4x\sqrt{-g}\left\{-\frac14F_{\mu\nu}F^{\mu\nu}+\frac{f}{2}\left[h''-(h')^2+h'\frac{f'}{f}\delta_{ki}+
\frac{2h'}{r}\delta_{0i}\right]A_iA^i\right\}.
\end{equation}
Here $i=0,1,2$; $k=1,2$. We can choose the gauge $A_r=0$ where the
second term is proportional to $A_\mu A^\mu$, i.e. it becomes the
genuine mass term. The theories~\eqref{13} and~\eqref{15} are
equivalent if the following surface terms are absent,
\begin{equation}
\label{16}
\int d^3x\left.\sqrt{-g}\,h'A_0^2\right|_{r=r_0}^{r=\infty}=0,
\end{equation}
\begin{equation}
\label{17}
\int d^3x\left. h'fA_k^2\right|_{r=r_0}^{r=\infty}=0.
\end{equation}
Now we can rewrite~\eqref{15} in the form of~\eqref{1} with some
potential $V(\psi)$ instead of the mass term $m^2\psi^2$. The
solution of equation of motion for $\psi$ in the absence of
$A_\mu$ must reproduce the mass term in~\eqref{15}, this dictates
the form of the potential $V(\psi)$.

The action~\eqref{15} leads to the equation of motion
\begin{equation}
\label{18}
A_x''+\frac{f'}{f}A_x'+\left[\frac{\omega^2}{f^2}+h''-(h')^2+h'\frac{f'}{f}\right]A_x=0.
\end{equation}
It is easy to check that the substitution $A_x\rightarrow e^h A_x$
converts Eq.~\eqref{14} to Eq.~\eqref{18}. The latter has the form
of Eq.~\eqref{7}. The term containing $\psi$ in Eq.~\eqref{7},
which was a numerical solution of coupled system of equations in
Ref.~\cite{HHH}, is replaced in Eq.~\eqref{18} by a term
completely dictated by the function $h$. We are free to choose
this function phenomenologically.

Consider the possibilities $h=r_0/r$ and $h=(r_0/r)^2$, where the
scaling~\eqref{11} is taken into account. They correspond to
solutions with condensates $\langle O_1\rangle$ and $\langle
O_2\rangle$ in~\eqref{5}. The resulting behavior of optical
conductivities $\sigma(\omega)$ is displayed in Figs.~6 and~7,
respectively. The obtained shapes recover (at least qualitatively)
the results of Ref.~\cite{HHH}. These two differing shapes are
expected for superconductors having the so-called type~II and
type~I coherence factors.

The considered possibilities can be generalized to the ansatz
\begin{equation}
\label{18b}
h=\alpha\left(\frac{r_0}{r}\right)^\beta.
\end{equation}
We may vary the parameter $\alpha$ at fixed $\beta$ and vice versa
and look at the behavior of optical conductivity. A couple of
corresponding examples are demonstrated in Figs.~8,~9 and 10,~11,
respectively. A non-polynomial ansatz for $h$ can be chosen. An
example is presented in Figs.~12 and~13.

The plots in Figs.~6-13 seem to share the following common
property: At any reasonable ansatz for $h$, when the form of
curves become maximally close to the experimental ones in Figs.~4
and~5, the minimal value of $\text{Im}[\sigma(\omega)]$ lies in
the vicinity of the region $\omega/r_0\simeq2$. According to our
discussions at the end of Section~2 and to relation~\eqref{4} for
$L=1$, $r_0\approx4.2T$, we have $\frac{\omega_g}{4.2T_c}\simeq2$,
i.e. we obtain a rough estimate for the gap coinciding\footnote{We
have tried many probe functions for $h$. A more accurate estimate
is $\omega/r_0\simeq1-3$ leading to the gap in the range
$\omega_g/T_c\simeq4-12$.} with~\eqref{10}.

This agreement looks surprising. It means that if a holographic
superconductor (in the probe limit) reproduces well the behavior
of the optical conductivity then the gap is approximately
universal (more strictly, of the same order of magnitude) for such
models due to the scaling symmetry. Even more surprising is that
the value of this gap is often close to that in the high-$T_c$
cuprates~\cite{gomes}.

The plots of $\text{Re}[\sigma(\omega)]$ in Fig.~8 shows a
formation of gap with changing $\alpha$. This suggests that in
reality $\alpha$ should be a decreasing function of $T$, say a
positive power of $(1-T/T_c)$. The parameter $\beta$ may also
include a temperature dependence with some critical value,
moreover, it might be responsible for the coherence factor type.

The main lesson of the exercises above is that given a concrete
superconducting material, in principle, one is able to find
phenomenologically the form of the dilaton background $h$ that
interpolates the observable superconducting properties of this
material. Then one can use the obtained $h$ to interpolate the
properties of other known high-$T_c$ superconductors and reveal
the required change of input parameters. This would help to
ascribe a physical meaning to the model parameters. Next step
would be a derivation of a phenomenologically reasonable form(s)
of $h$ from a consistent gauge/gravity theory. As usual, this step
is the most difficult.

In the probe limit, the form of the dilaton profile
$-2h\left(\frac{r}{r_0}\right)$ should follow from an action
\begin{equation}
\label{19}
S_\varphi=\int d^4x\sqrt{-g}\left(-\frac12\partial_\mu\varphi\partial^\mu\varphi-V(\varphi)\right),
\end{equation}
with the black hole metric~\eqref{2}. A solution of equation of
motion should yield $\langle\varphi\rangle\equiv h$. The potential
$V(\varphi)$ can be restored from the form of
$h\left(\frac{r}{r_0}\right)$. For instance, if we choose
$2h=\left(\frac{r}{r_0}\right)^\beta$ then
\begin{equation}
\label{20}
V=\beta\left[\frac12(\beta+3)\varphi^2-\frac{\beta^2r_0^3}{2-3/\beta}\varphi^{2-3/\beta}\right].
\end{equation}
Imposing the Breitenlohner--Freedman boundary for the scalar mass
in the AdS$_{d+1}$ space-time~\cite{BF},
$m^2\geq-\frac{d^2}{4L^2}$, we have $\beta(\beta+3)\geq-9/4$ that
leads to $(\beta-3/2)^2\geq0$, i.e. any value of $\beta$ is
allowed, with the value $\beta=-3/2$ providing the minimal mass.
The second term in~\eqref{20} looks unnatural since it emerges due
to the horizon and contains a non-integer power of interaction
except some special cases. This situation is quite general. If we
take the point of view that in the holographic approach, which is
inherently large-$N$ one, only quadratic in fields part is
relevant, then we should either assume $r_0=0$ in~\eqref{20}, i.e.
somehow motivate pure AdS metric in~\eqref{19}, or exclude the
interaction term. In the latter case, the equation of motion for
$\varphi=\varphi(r)$,
\begin{equation}
\label{21}
-\left(r(r^3-r_0^3)\varphi'\right)'+r^2\beta(\beta+3)\varphi=0,
\end{equation}
has solutions in terms of the Legendre functions of the first (P)
and second (Q) kind,
\begin{equation}
\label{22}
\varphi^{(1)}=P_{\beta/3}\left(1-\frac{2r^3}{r_0^3}\right),\qquad
\varphi^{(2)}=Q_{\beta/3}\left(1-\frac{2r^3}{r_0^3}\right).
\end{equation}
The energy functional of the action~\eqref{19} is minimal at
$\beta=-3/2$. Accepting this value and choosing\footnote{The
choice $h=P_{-1/2}(1-2(r/r_0)^3)$ does not satisfy the boundary
condition~\eqref{7b}.} $h=Q_{-1/2}(1-2(r/r_0)^3)$ in~\eqref{13},
we obtain a realistic prediction for the optical conductivity
which is depicted as one of plots in Figs.~14 and~15.

For comparison, we reproduce in Fig.~16 a certain experimental
plot for $\text{Re}[\sigma(\omega)]$ demonstrating the difference
of the shape below and above the critical temperature $T_c$. It is
seen that in reality the parameter $\beta$ in~\eqref{20} should
depend on $T$ achieving a critical value between $\beta/3=-0.7$
and $\beta/3=-0.75$ where the local minimum in $\text{Re}[\sigma]$
disappears. In contrast to the Landau--Ginzburg theory, the
quadratic part of the potential~\eqref{20} does not vanish at the
critical point. This feature could have a natural interpretation
--- the high-$T_c$ superconductors represent doped materials and their
$T_c$ depends crucially on the density of doped holes $x$ that
lies in a finite interval $a<x<b$~\cite{carlson}. The maximal
$T_c$ roughly corresponds to $x=(a+b)/2$ (the optimal doping).
Thus the parameter $\beta$ in~\eqref{20} might encode a dependence
on the doping $x$.

\section{Discussions}

The presented approach to building a bottom-up holographic
superconductor is more phenomenological than the original
one~\cite{HHH}. It would be interesting to clarify a possible
origin of our approach. The SW holographic model in hadron physics
cannot help because its relation to any fundamental string theory
is unknown. The full action for the considered holographic
superconductor is
\begin{equation}
\label{24}
S=\kappa_1 S_{\text{grav}}+\kappa_2 S_{\varphi}+\kappa_3 S_{\varphi,A}.
\end{equation}
Our approach is justified if
\begin{equation}
\label{25}
\kappa_1\gg\kappa_2\gg\kappa_3.
\end{equation}
By assumption, the part $\kappa_1 S_{\text{grav}}$ gives the black
hole metric~\eqref{2} and the backreaction to the metric from
$\kappa_2 S_{\varphi}$ and $\kappa_3 S_{\varphi,A}$ can be
neglected due to~\eqref{25}. After that we find
$\langle\varphi\rangle$ from minimum of $S_{\varphi}$ and then
extract the phenomenology from $S_{\varphi,A}$. Following the
analogy with the SW model in hadron physics~\cite{son2}, we have
chosen the interaction of the scalar field $\varphi$ with the
gauge field $A$ in the form $e^{\varphi}F_{\mu\nu}^2$. In
principle, such coupling is provided by some known dilatonic black
holes unstable to forming the scalar hair~\cite{horowitz}.
However, this possibility has not been exploited in holographic
superconductors because $F_{\mu\nu}^2$ acts as a source for
$\varphi$, so $\langle\varphi\rangle\neq0$ for any charged black
hole~\cite{horowitz}. In our case, the given argument does not
apply since, first, $\varphi$ in the exponent represents a fixed
background, second, we are free to choose a different coupling,
say, $e^{\varphi^2}F_{\mu\nu}^2$.

How the action~\eqref{24} with the sequence of scales~\eqref{25}
may follow from a string theory is an open problem. In the case of
holographic dual for a conformal $SU(N)$ gauge theory, one has
$\kappa_1=\mathcal{O}(N^2)$. We may implement~\eqref{25} assuming
the scaling
\begin{equation}
\label{26}
\kappa_1=\mathcal{O}(N^2),\qquad \kappa_2=\mathcal{O}(N),\qquad \kappa_3=\mathcal{O}(1),
\end{equation}
at large $N$. This scaling entails the question about the physical
meaning of $N$ in the holographic superconductors. The given
question is tightly related with a long standing problem
concerning the analog of large-$N$ limit in condensed matter
systems~\cite{horowitz}.

Perhaps the answer lies in the structure of high-$T_c$
superconductors. They represent small pieces of layered
materials~\cite{gomes}. Imagine that $N$ is the number of layers.
Imagine further that each layer interacts with all (or a
substantial part of) other layers. Then we have $\mathcal{O}(N^2)$
interactions and the corresponding background medium is described
in the action~\eqref{24} by the part $\kappa_1 S_{\text{grav}}$
with $\kappa_1=\mathcal{O}(N^2)$. Since in the experimental
patterns the layers are weakly coupled~\cite{horowitz}, we should
have a weakly coupled gravitational theory as it is required by
the AdS/CFT correspondence. Assume that the medium inside each
layer can be modelled by a relativistic theory of the scalar field
$\varphi$. Since we have $N$ layers, in~\eqref{24} the
contribution $S_{\varphi}$ emerges with $\kappa_2=\mathcal{O}(N)$.
The real high-$T_c$ materials can have several sorts of layers
composed of different atoms. So, strictly speaking, we should
consider several fields $\varphi_i$. But we simplify the
situation. The part $\kappa_3 S_{\varphi,A}$ describes a response
of the system to the electromagnetic field. This response is
unrelated to the number of layers, thus $\kappa_3=\mathcal{O}(1)$.

The outlined physical picture could justify the choice of the
action~\eqref{24} with the behavior of constants~\eqref{26} at
large $N$. In some sense, within this interpretation a layered
material resembles the stack of $N$ branes in the AdS/CFT
correspondence~\cite{aharony}. Such a resemblance might be among
reasons why the gauge/string duality can be successfully applied
to the high-$T_c$ superconductors.

In our approach, the dilaton profile comes from condensation of a
scalar field. This description should be dual to a theory of
S-wave superconductors. Many real high-$T_c$ superconductors are
known to have the $D$-wave (the cuprates) or sometimes $P$-wave
superconductivity~\cite{gomes}. The construction of the
corresponding extensions is straightforward
--- one should replace the contribution $S_{\varphi}$ in~\eqref{24}
by an action for the vector or tensor field that condenses and
forms the dilaton profile in $S_{\varphi,A}$. In principle, the
vector or tensor condensate can be made anisotropic and, hence,
describe the anisotropic (striped) superconductivity.

\section{Conclusions}

A fruitful interrelation between particle physics and
superconductivity in condensed matter started in early 60-th when
Nambu and Jona-Lasinio proposed their famous model for the
nucleons from an analogy with the theory of
superconductivity~\cite{njl}. Recently the story has taken an
interesting turn with the appearance of the holographic approach
inspired by the gauge/gravity duality in string theory. It turned
out that the concept of the bottom-up holographic models
originally developed for the hadron physics can be applied to the
high-$T_c$ superconductors~\cite{HHH}, the nature of which remains
enigmatic. Unfortunately the existing holographic superconductors
are still very far from any quantitative description of the real
high-$T_c$ materials. In this respect, the progress in the
holographic hadron physics is more considerable.

In the present work, we have made an attempt to advance in the
quantitative holographic description of the high-$T_c$
superconductors. For this purpose we applied the idea of the soft
wall holographic model from hadron physics~\cite{son2}. The
phenomenological description of the form of the optical
conductivity in our approach is in one-to-one correspondence with
the phenomenological description of the form of the hadron
spectrum in the standard SW model~\cite{son2}. The simplest model
of holographic superconductor is believed to be the one proposed
in the original paper~\cite{HHH}. We have shown that it is
possible to construct much simpler models. The simplicity is
likely the main advantage of the proposed SW holographic
superconductors. In hadron physics, the SW model does not describe
(without significant complications) the dynamics of the chiral
symmetry breaking but in description of the observable spectrum it
seems to be the most successful bottom-up holographic approach. In
the holographic superconductors, we have a similar situation: The
considered SW model (in its current form) does not describe the
phase transition to superconductivity, but if we are interested in
the experimentally measurable quantities like optical
conductivity, the SW superconductors are able to describe the
emerging phenomenology more realistically than many other
holographic superconductors.

There are various directions for extensions and applications of
the proposed approach. We mention some of them. (i) It would be
interesting to identify the most reasonable dilaton profiles from
the best fit to the experimental data on the optical
conductivities. (ii) One can calculate other transport properties
and study the response to the non-zero chemical potential and
magnetic field. (iii) At some choices of the dilaton profile the
SW model seems to describe a superinsulator and this could be
interesting. (iv) Perhaps the phase transition to
superconductivity can be modelled as the Hawking-Page phase
transition in the gravitational part of~\eqref{24} from the
thermal AdS space to AdS with black hole~\cite{Hawking}. The
transition point could relate the critical temperature to some
dimensional model parameter (as in Ref.~\cite{Herzog:2006} for the
deconfinement temperature). The requirement to describe the
transport properties of the high-$T_c$ materials above $T_c$ may
then restrict significantly a possible form of the dilaton
profile. (v) Among theoretical challenges for the holographic
superconductors one can mention the problem of a dual
gravitational interpretation for the empirical Homes' scaling law
in the high-$T_c$ superconductors~\cite{homes}. Our preliminary
analysis showed that this law imposes rather strong constraint on
the mutual dependence of the model parameters and seems to
prescribe for them a certain temperature dependence. We hope to
address at least some of the issues above in the future work.

\section*{Acknowledgments}

The authors acknowledge Saint-Petersburg State University for
research grants 11.38.189.2014 and 11.38.660.2013. The work was
also partially supported by the RFBR grant 13-02-00127-a.

%\newpage

\section*{Appendix: Plots}

%\begin{figure}[h]
%\centering
%\includegraphics[width=0.5\textwidth]{fig1.eps}
% \includegraphics[width=0.5\textwidth]{fig1c.eps}
%\caption{The condensate $\langle O_2 \rangle$ as a function of
%temperature (see text).}
%\end{figure}

\begin{figure}[ht]
\centering
\begin{minipage}[ht]{0.7\linewidth}
\includegraphics[width=0.8\linewidth]{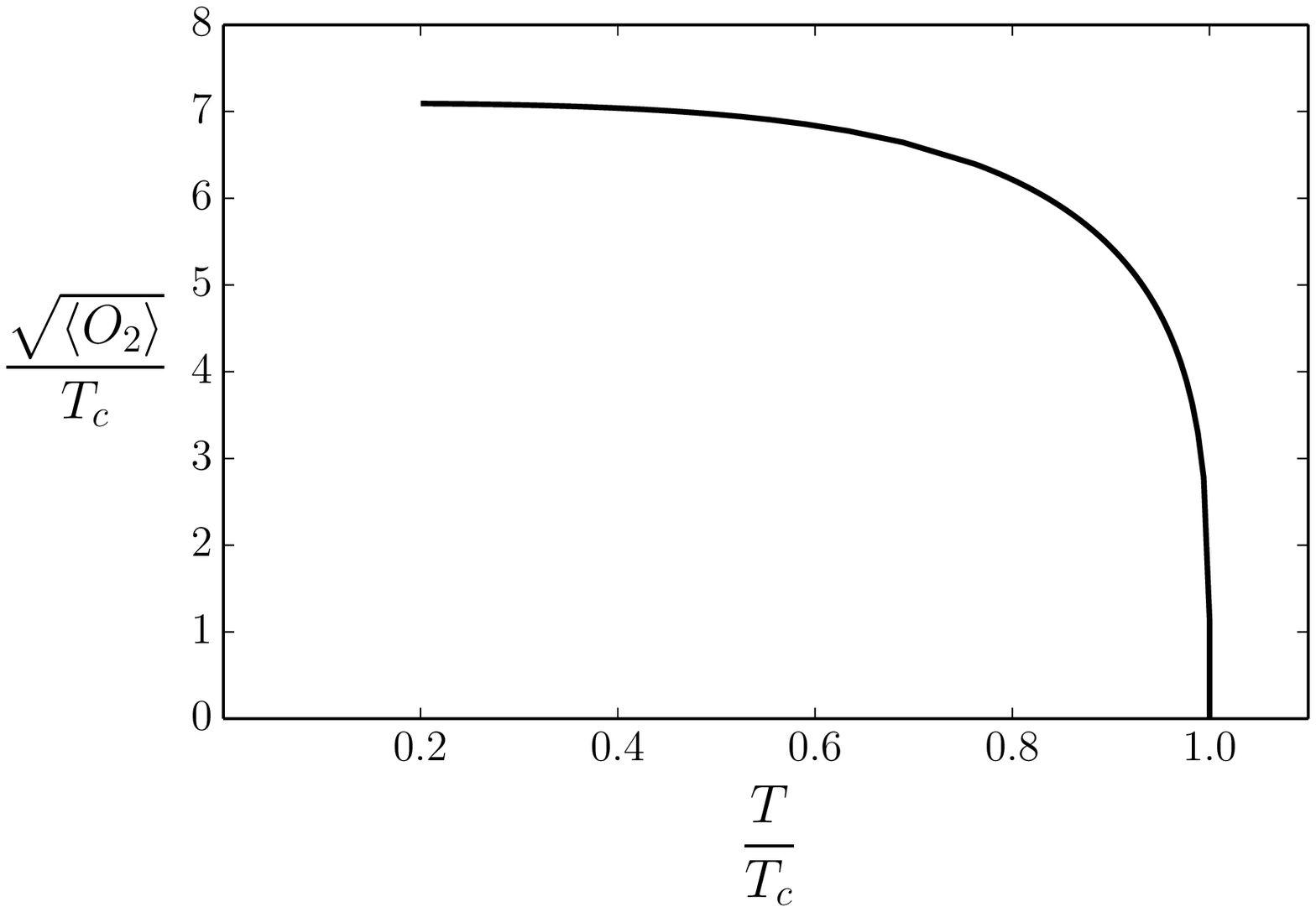} \\
{\scriptsize Fig. 1. The condensate $\langle O_2 \rangle$ as a
function of temperature for $m^2=-2$~\cite{cai}.}
\end{minipage}
\end{figure}

\begin{figure}[ht]
\begin{minipage}[ht]{0.46\linewidth}
\includegraphics[width=1\linewidth]{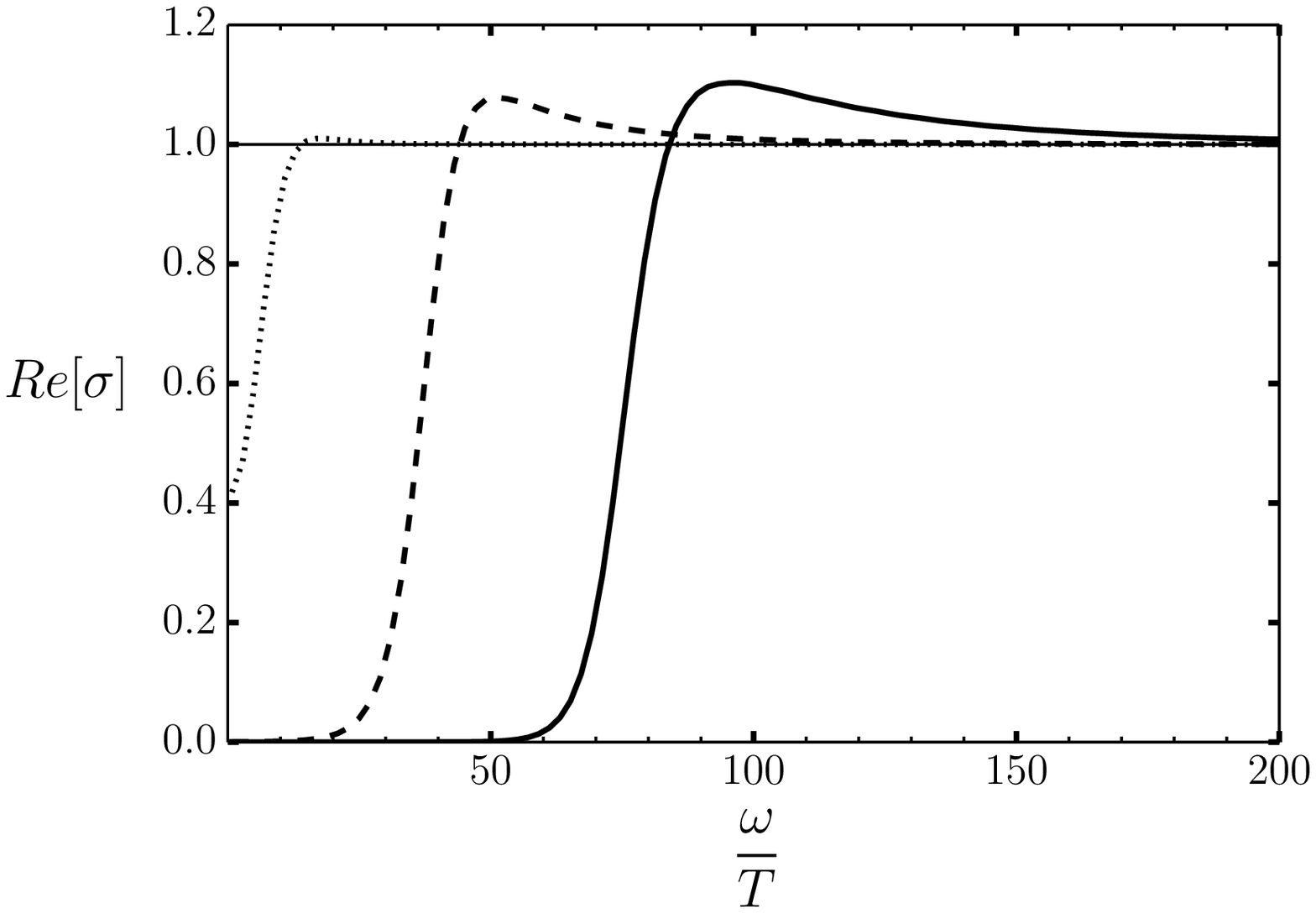} \\
{\scriptsize Fig. 2. The real part of the optical conductivity as
a function of frequency normalized by temperature for $m^2=-2$.
Curves from left to right correspond to
$\frac{T}{T_c}=\lambda\approx 0.888$ (dotted), $\lambda\approx
0.222$ (dashed) and $\lambda\approx 0.105$ (solid)~\cite{cai}.}
\end{minipage}
    \hfill
   \begin{minipage}[ht]{0.46\linewidth}
    \includegraphics[width=1\linewidth]{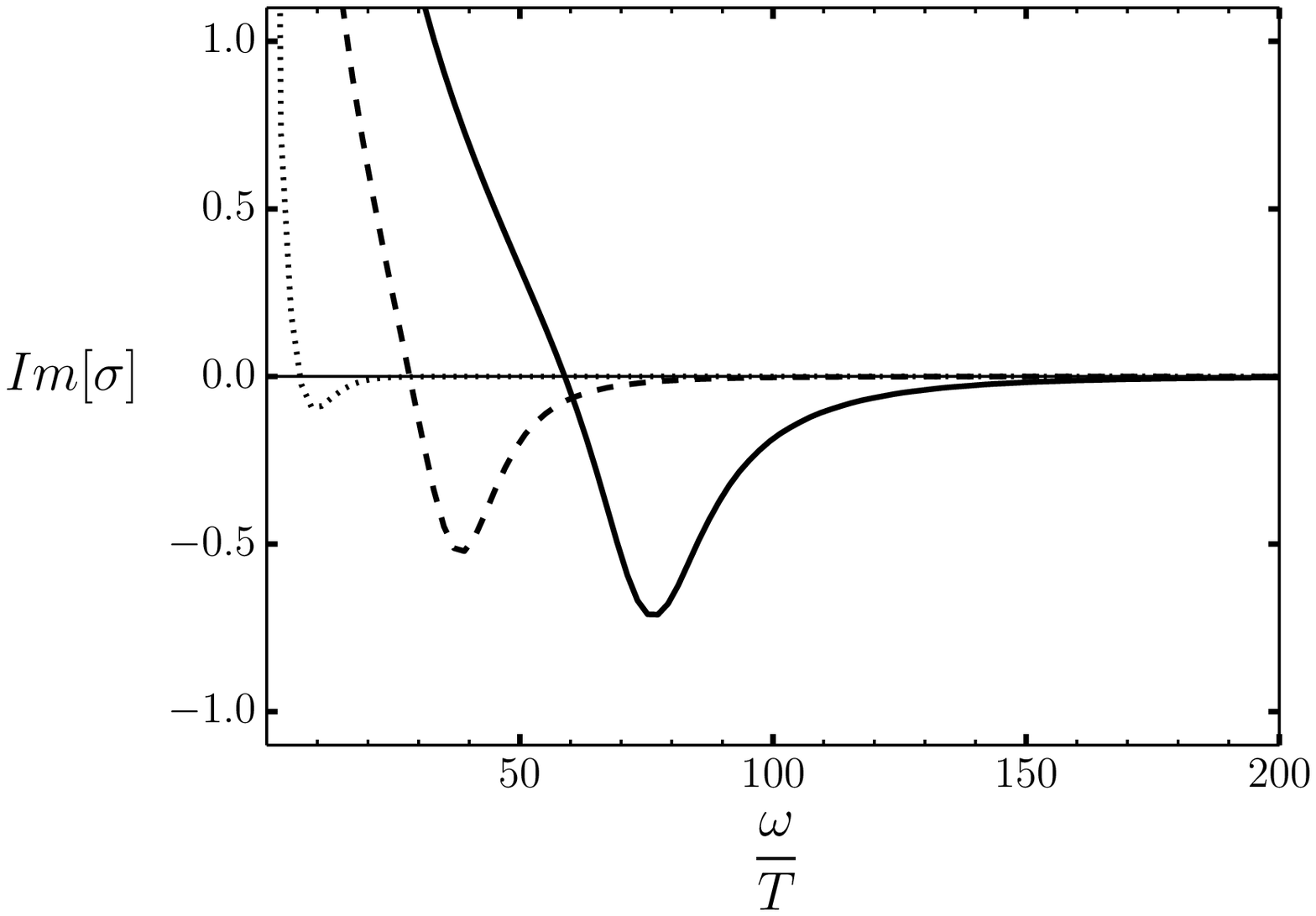} \\
    {\scriptsize Fig. 3. The imaginary part of the optical conductivity from Fig.~2~\cite{cai}.\vspace{1.6cm} }
    \end{minipage}
\end{figure}

\begin{figure}[ht]
\begin{minipage}[ht]{0.46\linewidth}
\includegraphics[width=1\linewidth]{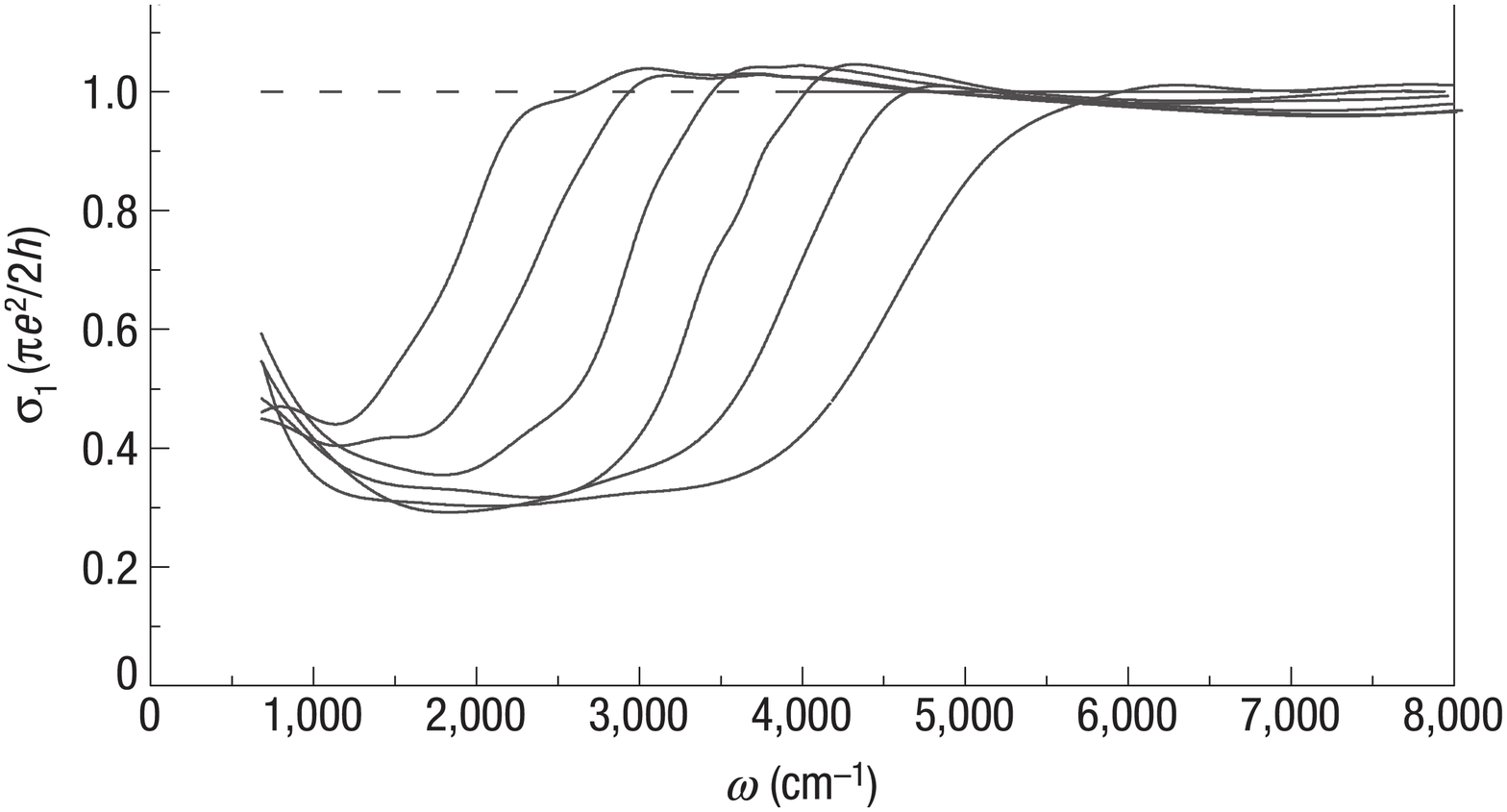} \\
{\scriptsize Fig. 4. The optical conductivity measured in
graphene~\cite{Li}. The curves from left to right correspond to
the gate voltage 0, 10, 17, 28, 40, 54, 71 V.}
\end{minipage}
    \hfill
   \begin{minipage}[ht]{0.46\linewidth}
    \includegraphics[width=1\linewidth]{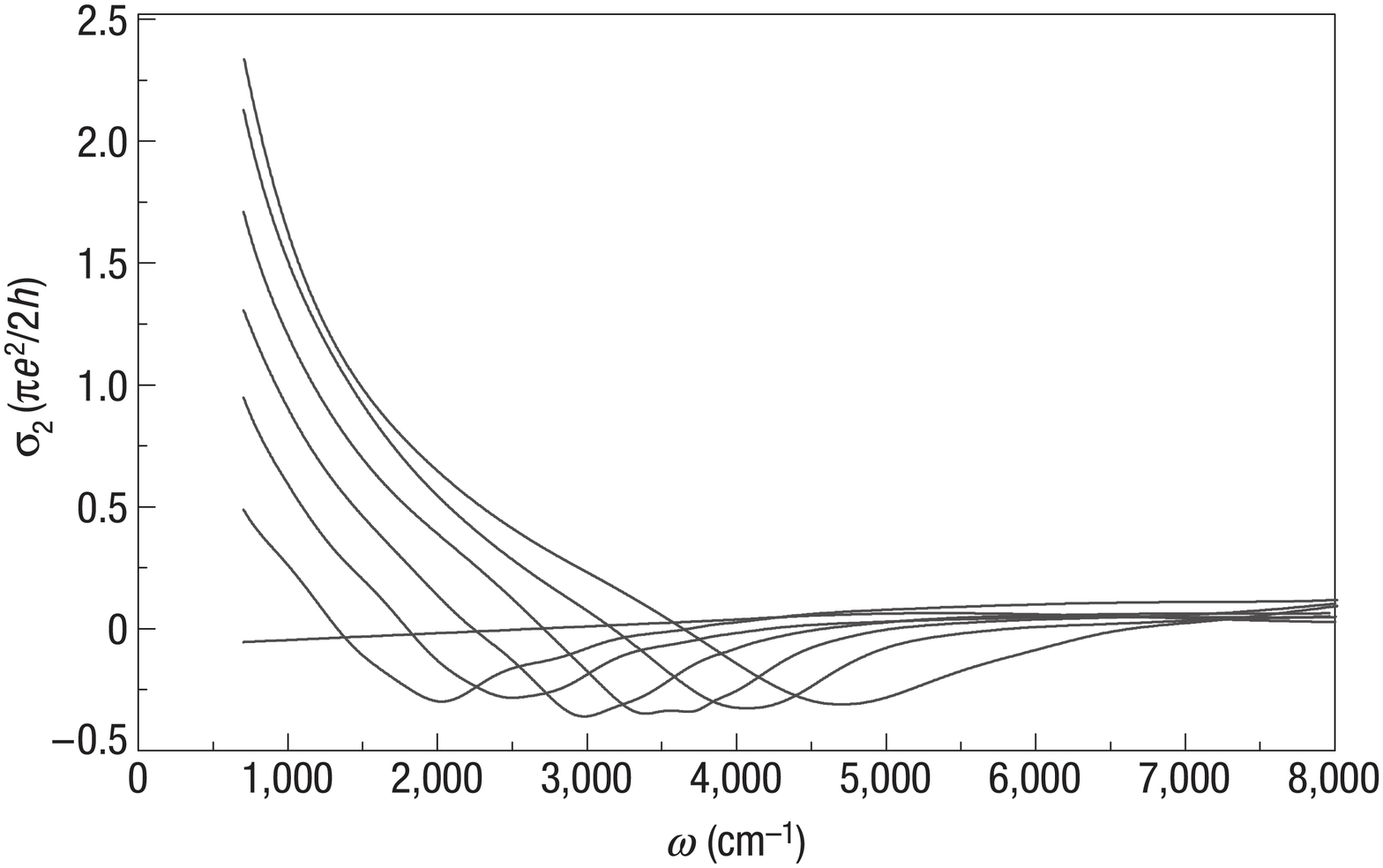} \\
    {\scriptsize Fig. 5. The imaginary part of the conductivity from Fig.~4~\cite{Li}.\vspace{0.6cm}}
    \end{minipage}
\end{figure}

\begin{figure}[ht]
\begin{minipage}[ht]{0.46\linewidth}
\includegraphics[width=1\linewidth]{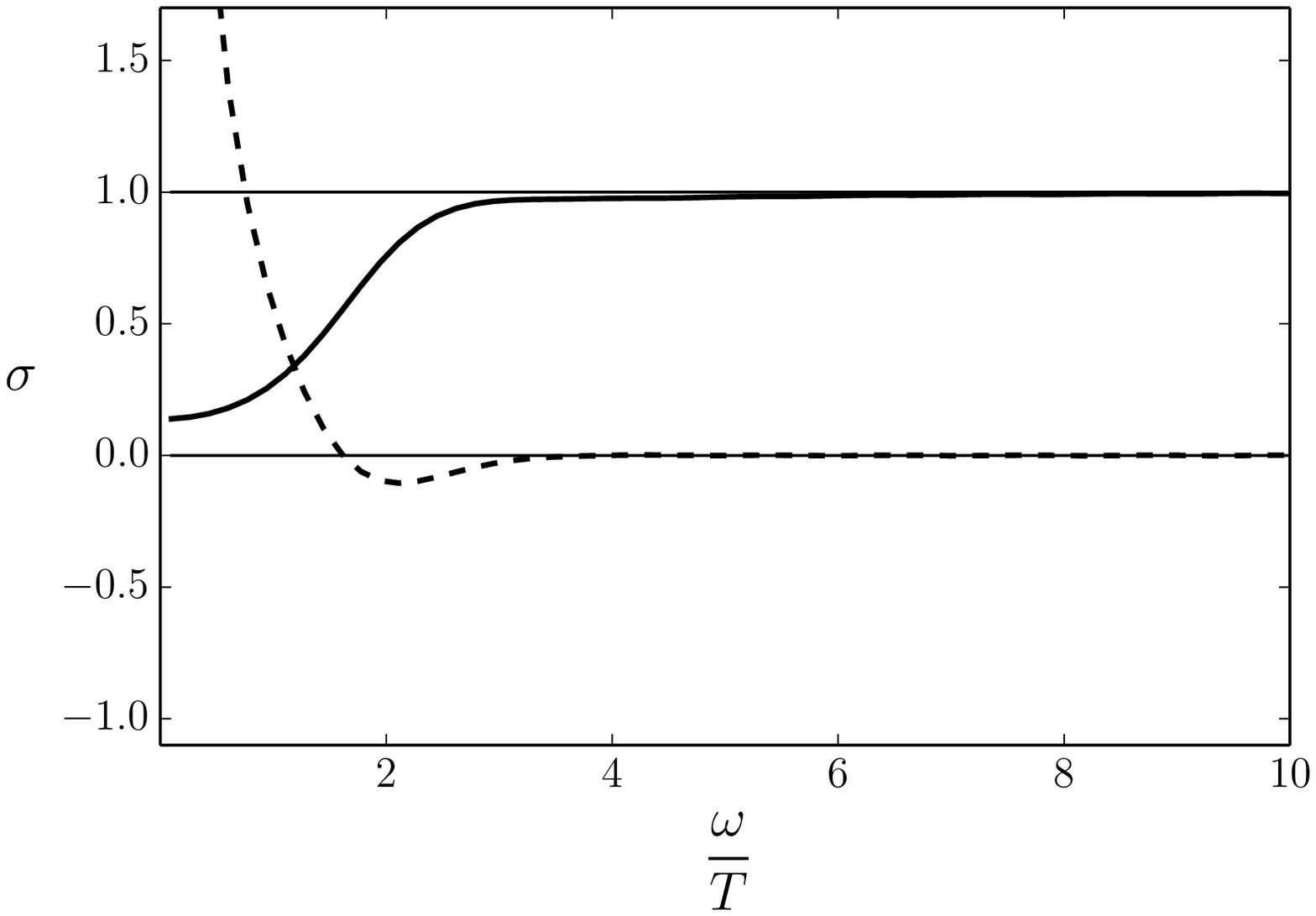} \\
{\scriptsize Fig. 6. The optical conductivity at $h=r_0/r$. The
solid line corresponds to $\text{Re}[\sigma]$ and the dashed one
to $\text{Im}[\sigma]$.}
\end{minipage}
    \hfill
   \begin{minipage}[ht]{0.46\linewidth}
    \includegraphics[width=1\linewidth]{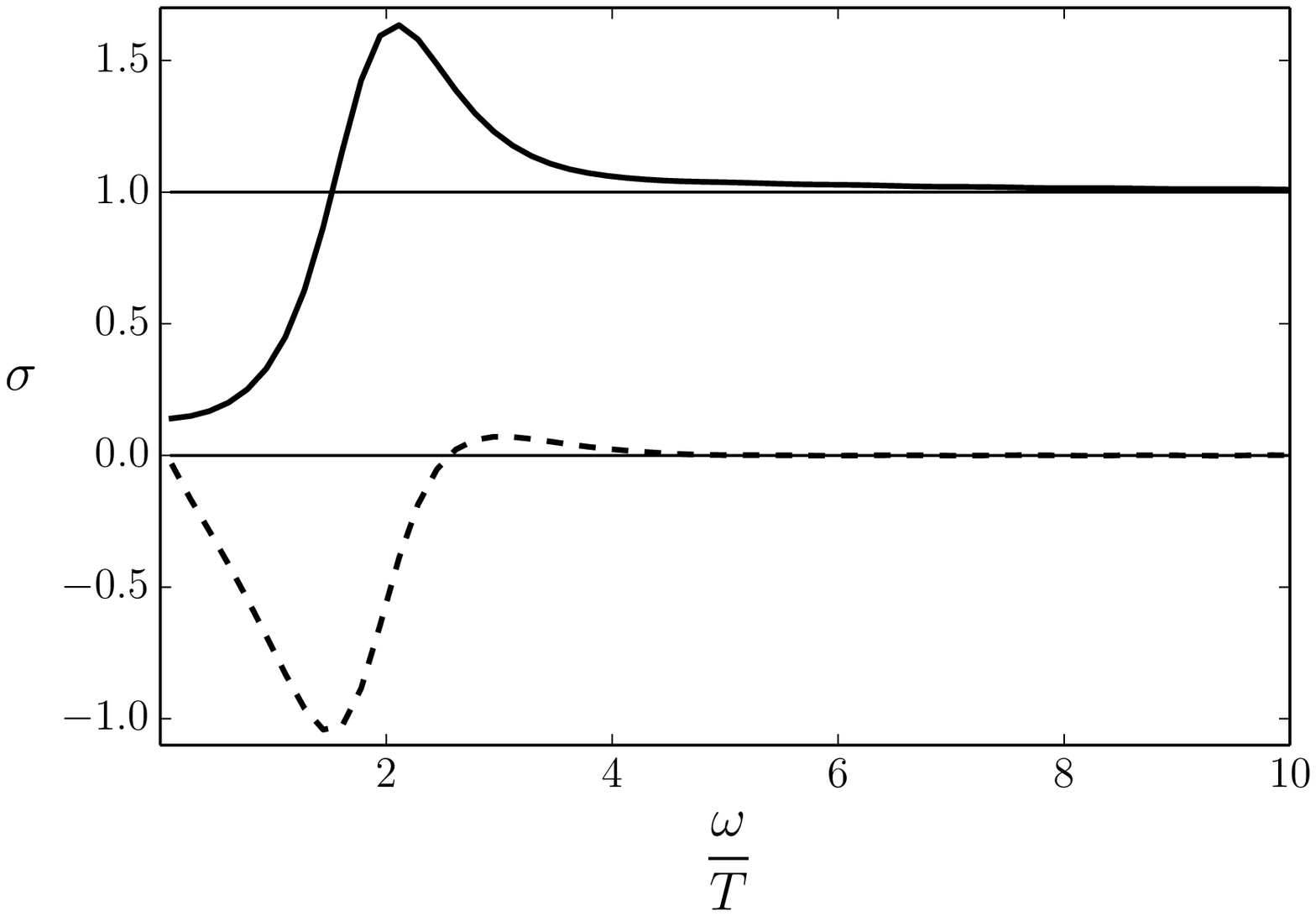} \\
    {\scriptsize Fig. 7. The optical conductivity at $h=(r_0/r)^2$.
    The notations are as in Fig.~6.\vspace{0.6cm}}
    \end{minipage}
\end{figure}

\begin{figure}[ht]
\begin{minipage}[ht]{0.46\linewidth}
\includegraphics[width=1\linewidth]{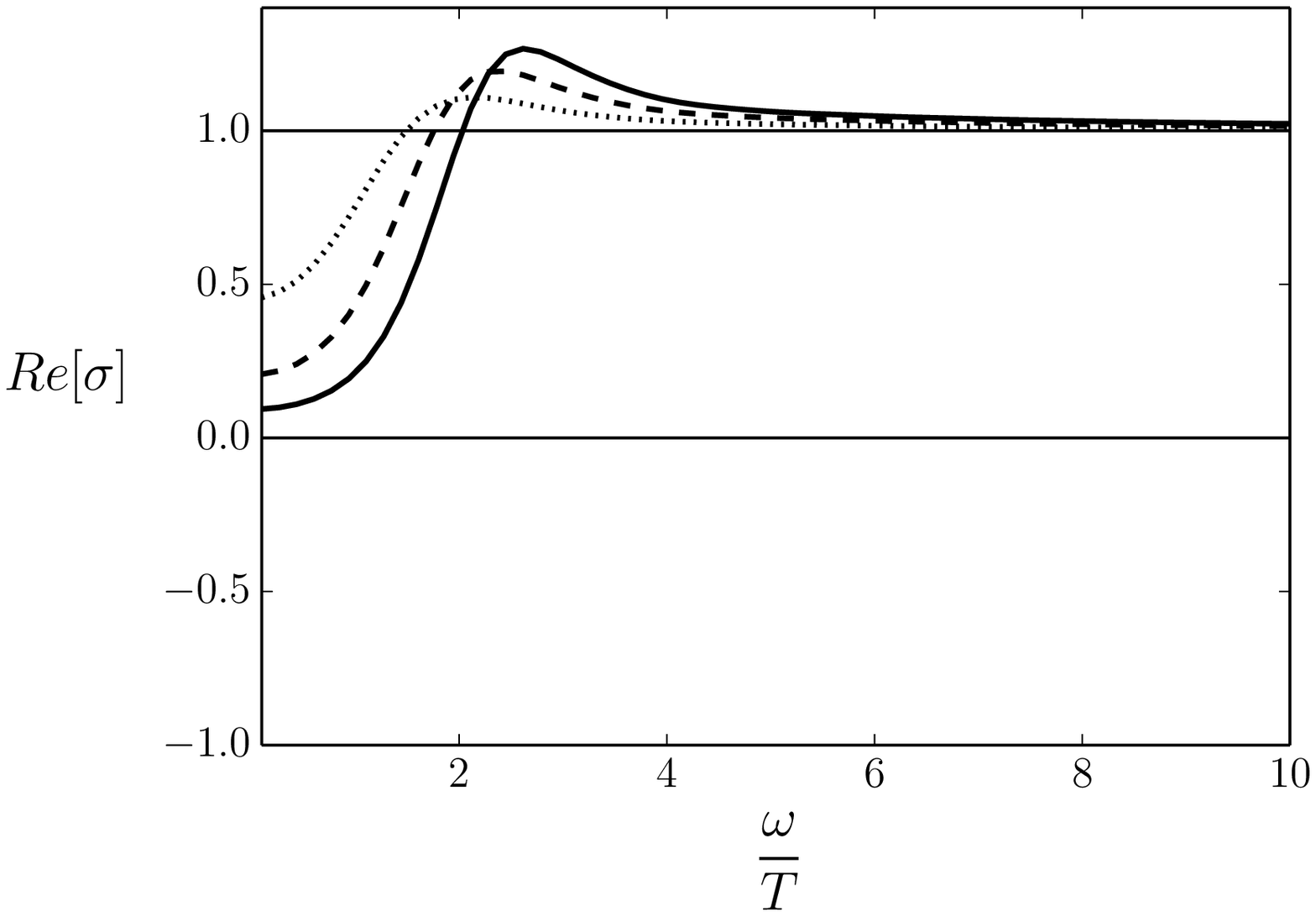} \\
{\scriptsize Fig. 8. $\text{Re}[\sigma]$ at
$h=\alpha\left(\frac{r_0}{r}\right)^{1.4}$. From left to right
$\alpha=0.4$ (dotted), $\alpha=0.8$ (dashed) and $\alpha=1.2$
(solid).}
\end{minipage}
    \hfill
   \begin{minipage}[ht]{0.46\linewidth}
    \includegraphics[width=1\linewidth]{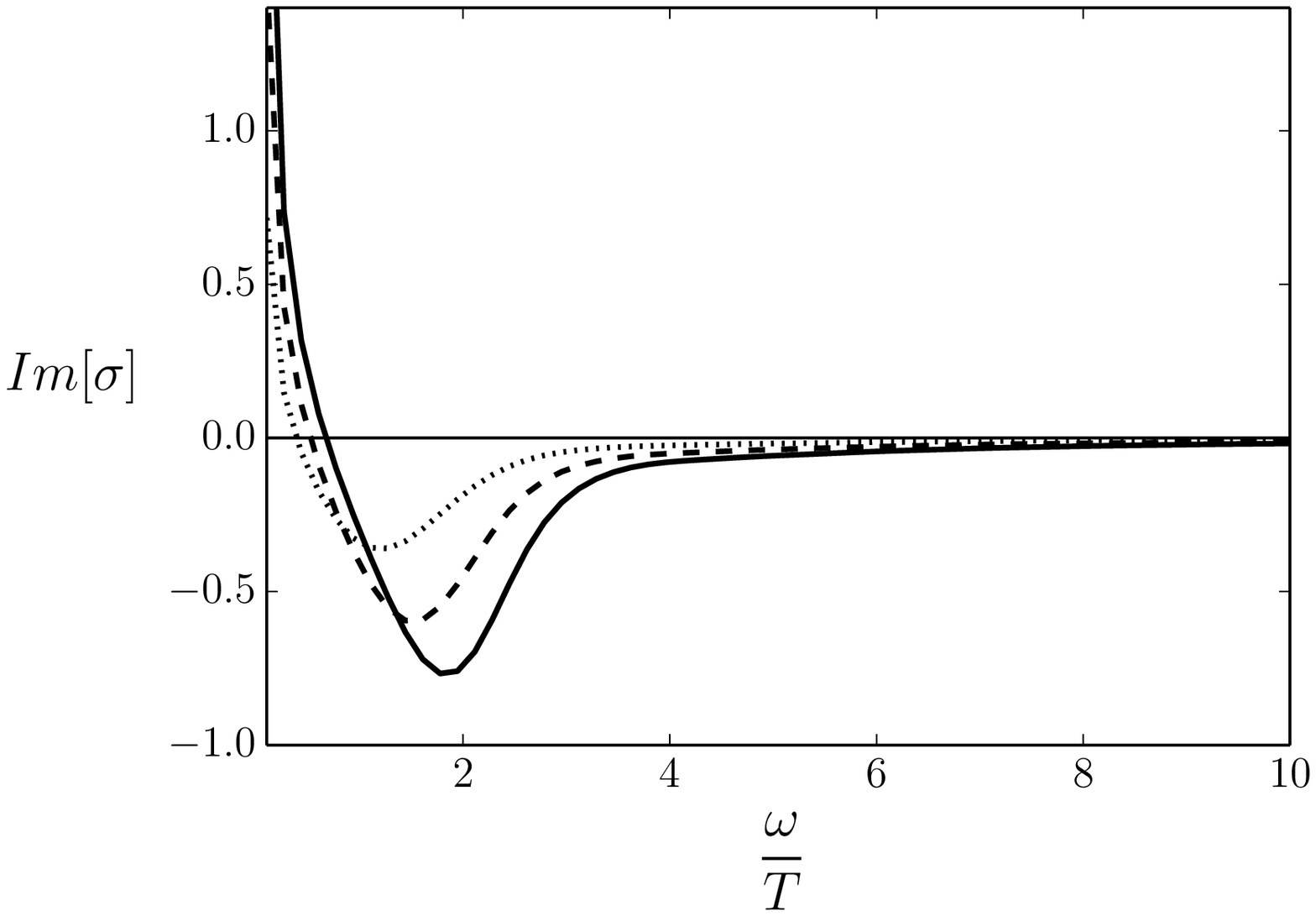} \\
    {\scriptsize Fig. 9. $\text{Im}[\sigma]$ corresponding to the plots in Fig.~8.\vspace{0.6cm}}
    \end{minipage}
\end{figure}

\begin{figure}[ht]
\begin{minipage}[ht]{0.46\linewidth}
\includegraphics[width=1\linewidth]{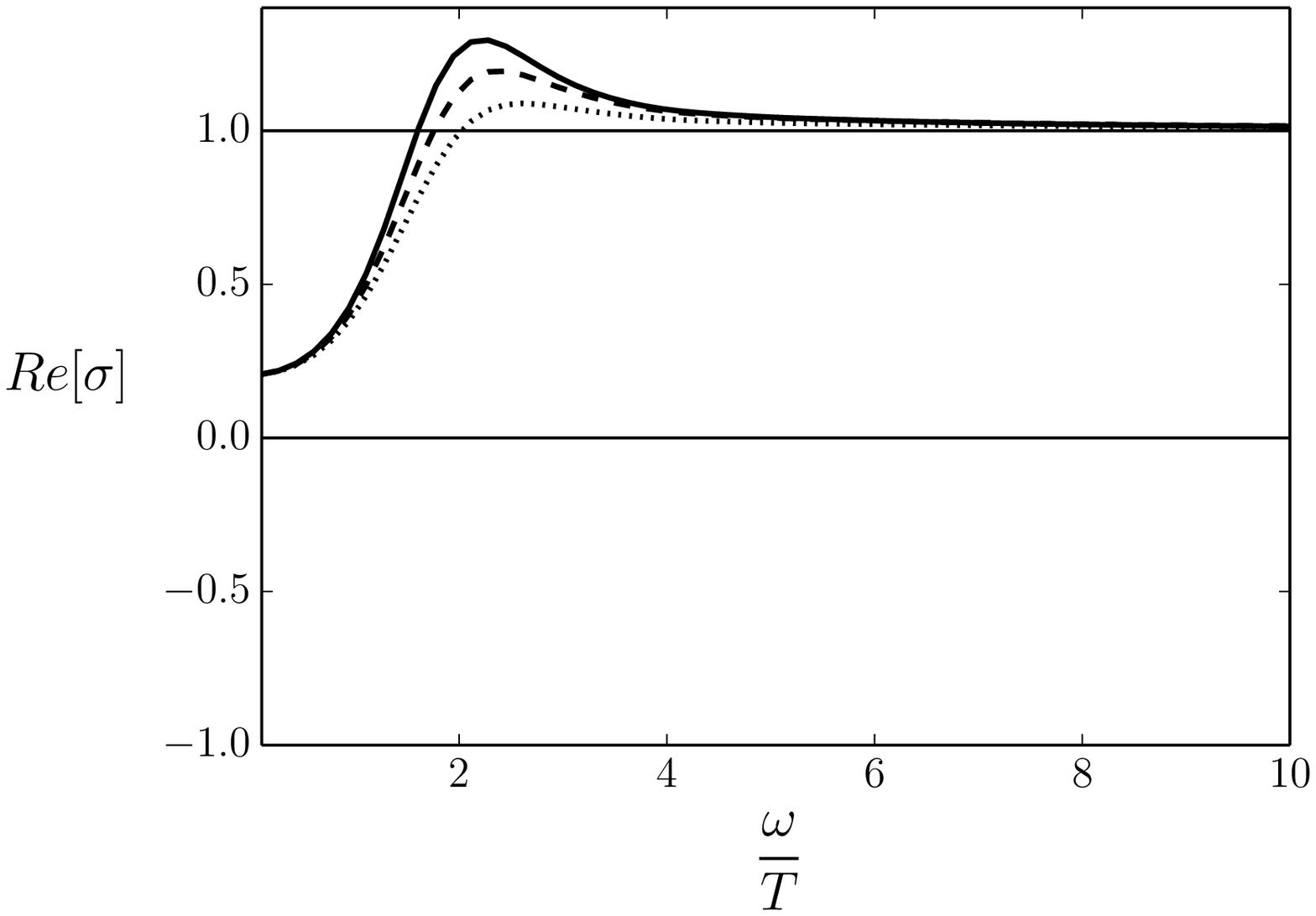} \\
{\scriptsize Fig. 10. $\text{Re}[\sigma]$ at
$h=0.8\left(\frac{r_0}{r}\right)^\beta$. From right to left
$\beta=1.2$ (dotted), $\beta=1.4$ (dashed) and $\beta=1.6$
(solid).}
\end{minipage}
    \hfill
   \begin{minipage}[ht]{0.46\linewidth}
    \includegraphics[width=1\linewidth]{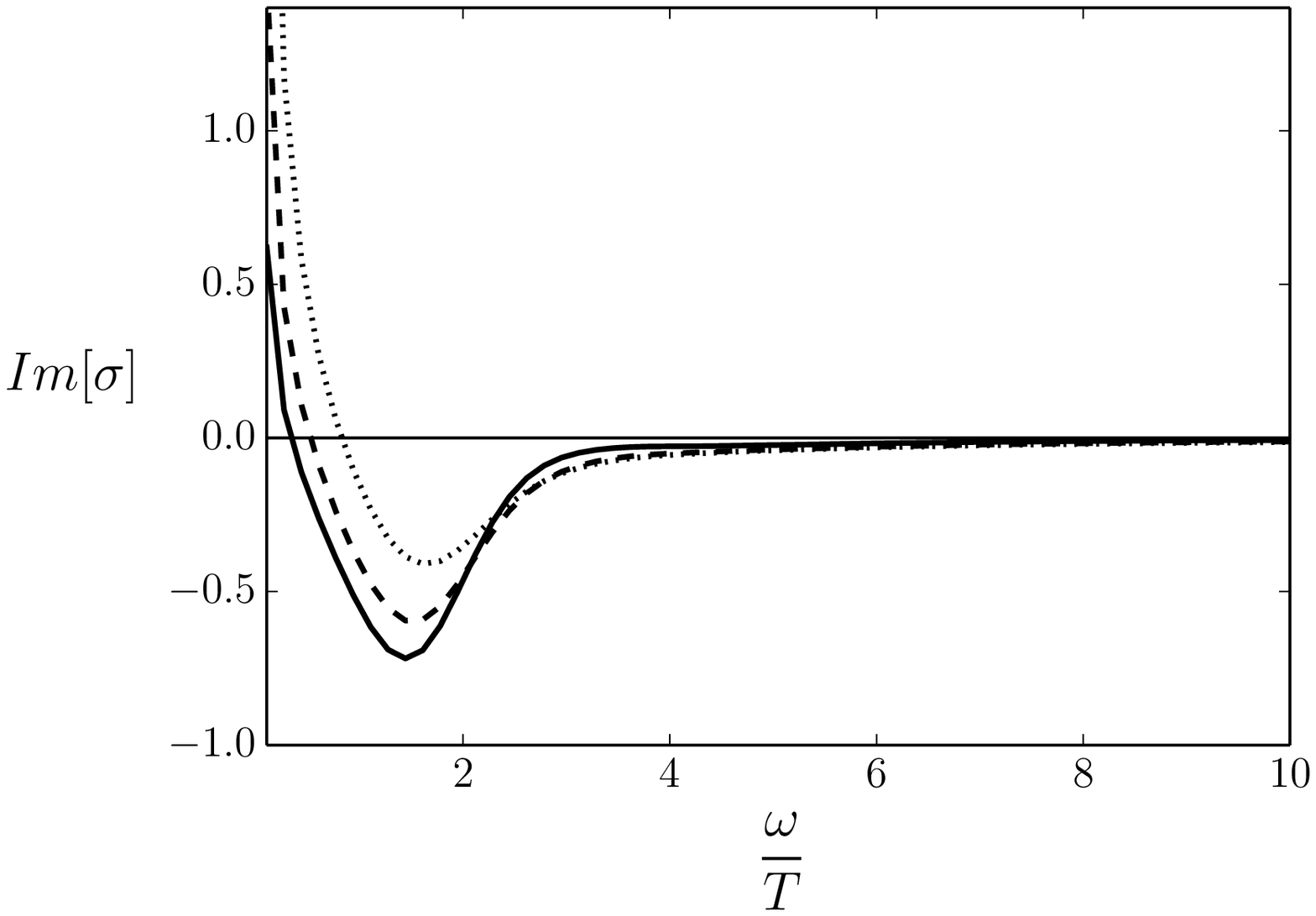} \\
    {\scriptsize Fig. 11. $\text{Im}[\sigma]$ corresponding to the plots in Fig.~10.\vspace{0.6cm}}
    \end{minipage}
\end{figure}

\begin{figure}[ht]
\begin{minipage}[ht]{0.46\linewidth}
\includegraphics[width=1\linewidth]{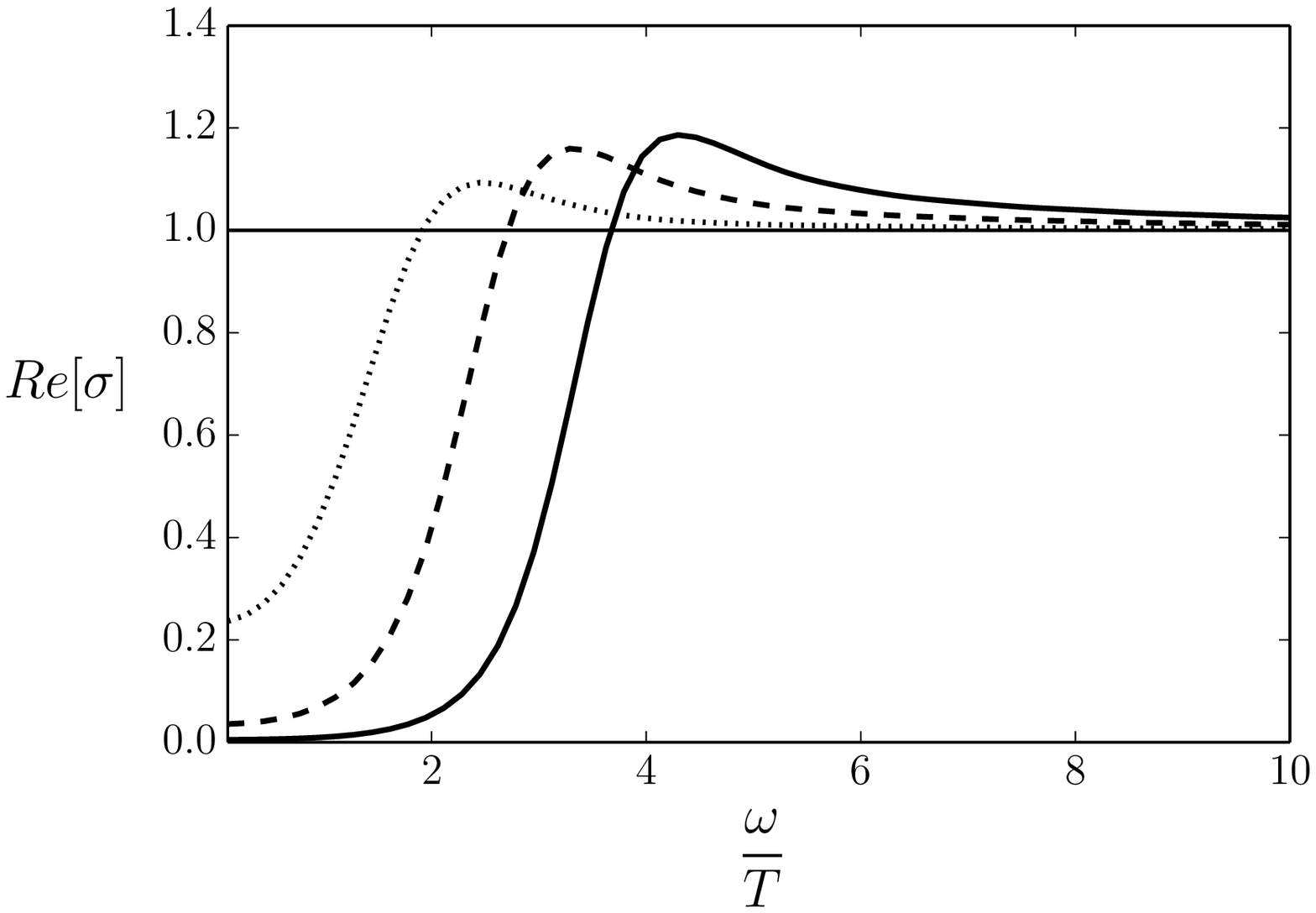} \\
{\scriptsize Fig. 12. $\text{Re}[\sigma]$ at
$h=\log\left(\cosh\left(\frac12+\alpha\frac{r_0}{r}\right)\right)$.
From left to right we take $\alpha=1$ (dotted), $\alpha=2$
(dashed) and $\alpha=3$ (solid).}
\end{minipage}
    \hfill
   \begin{minipage}[ht]{0.46\linewidth}
    \includegraphics[width=1\linewidth]{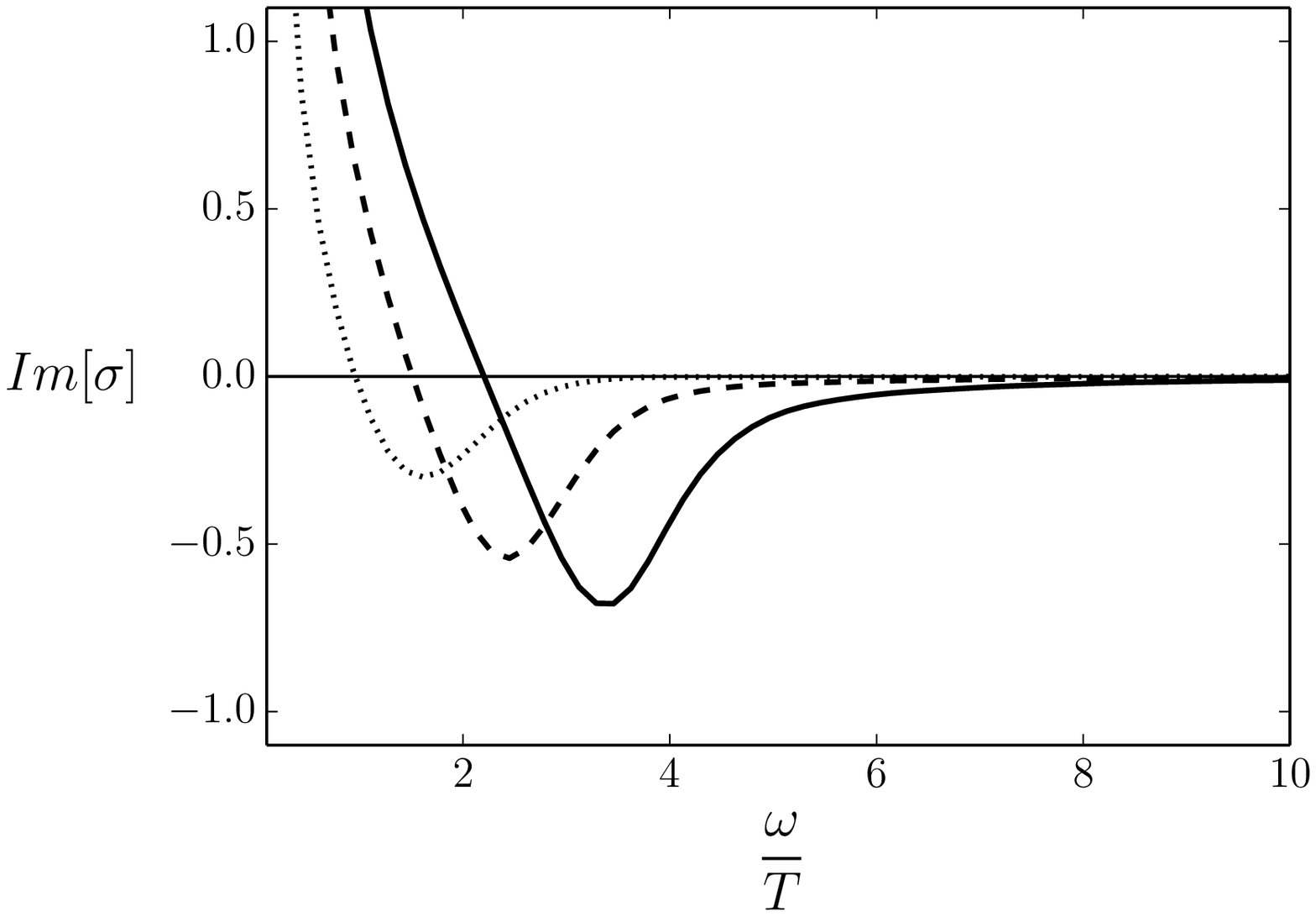} \\
    {\scriptsize Fig. 13. $\text{Im}[\sigma]$ corresponding to the plots in Fig.~12.\vspace{0.6cm}}
    \end{minipage}
\end{figure}

\begin{figure}[ht]
\begin{minipage}[ht]{0.46\linewidth}
\includegraphics[width=1\linewidth]{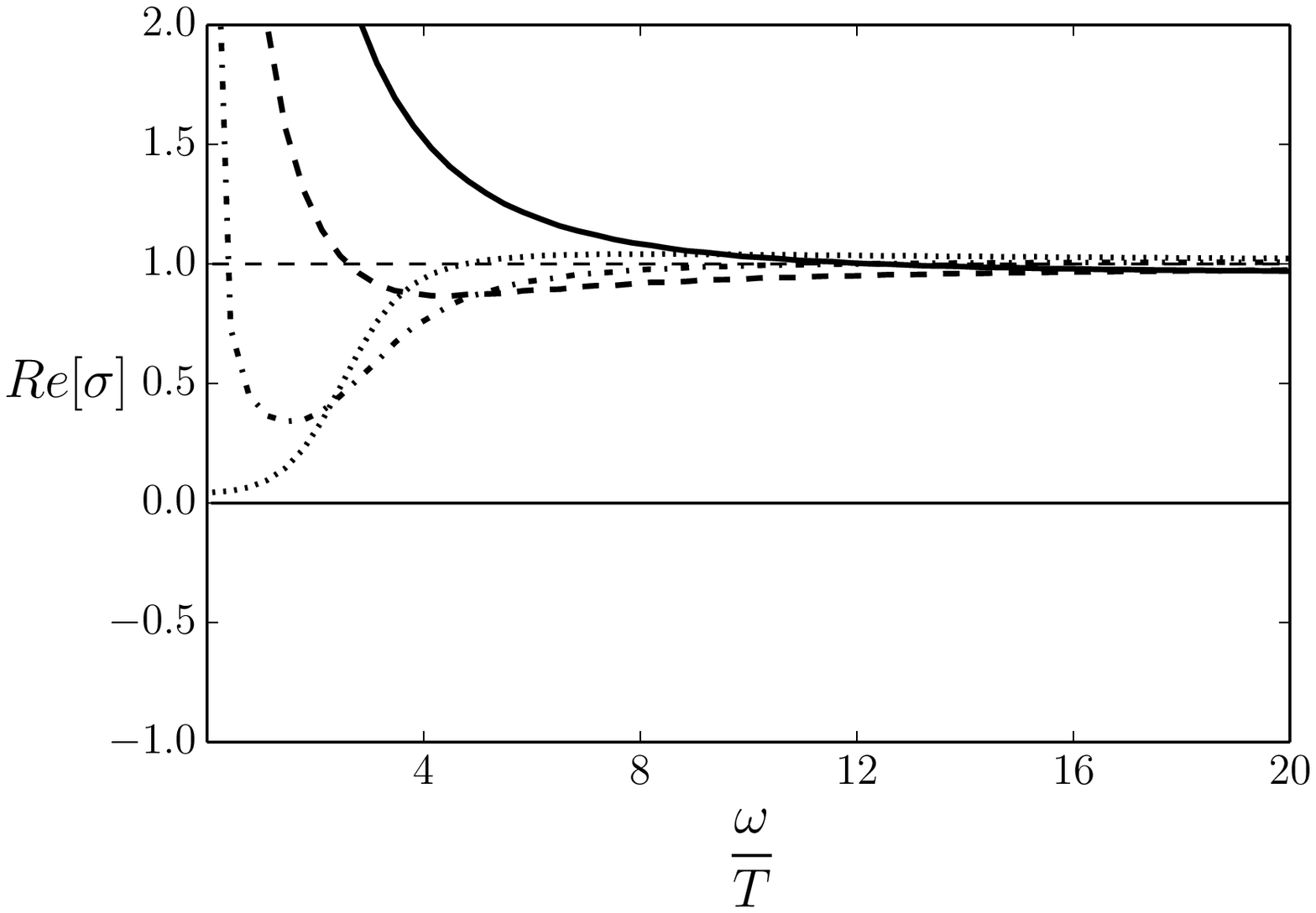} \\
{\scriptsize Fig. 14. $\text{Re}[\sigma]$ at
$h=Q_{\nu}(1-2(r/r_0)^3)$. From left to right $\nu\!=\!-0.5$ (with
an approximate $\delta$-function at $\frac{\omega}{T}\!=\!0$),
$\nu\!=\!-0.6$, $\nu\!=\!-0.7$, $\nu\!=\!-0.75$.}
\end{minipage}
    \hfill
   \begin{minipage}[ht]{0.46\linewidth}
    \includegraphics[width=1\linewidth]{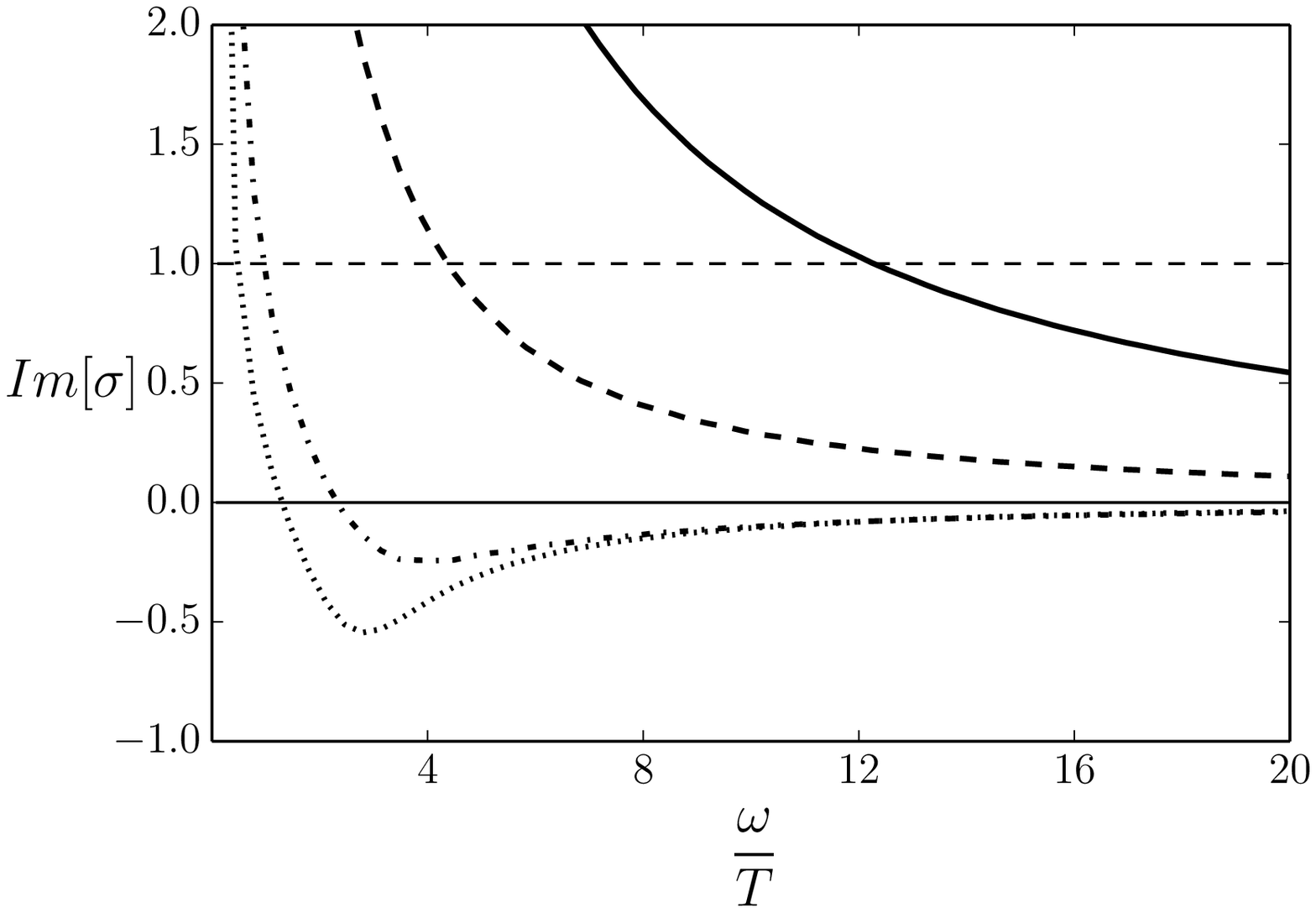} \\
    {\scriptsize Fig. 15. $\text{Im}[\sigma]$ corresponding to the plots in Fig.~14.\vspace{0.6cm}}
    \end{minipage}
\end{figure}

\begin{figure}[ht]
\centering
\begin{minipage}[ht]{0.6\linewidth}
\includegraphics[width=1\linewidth]{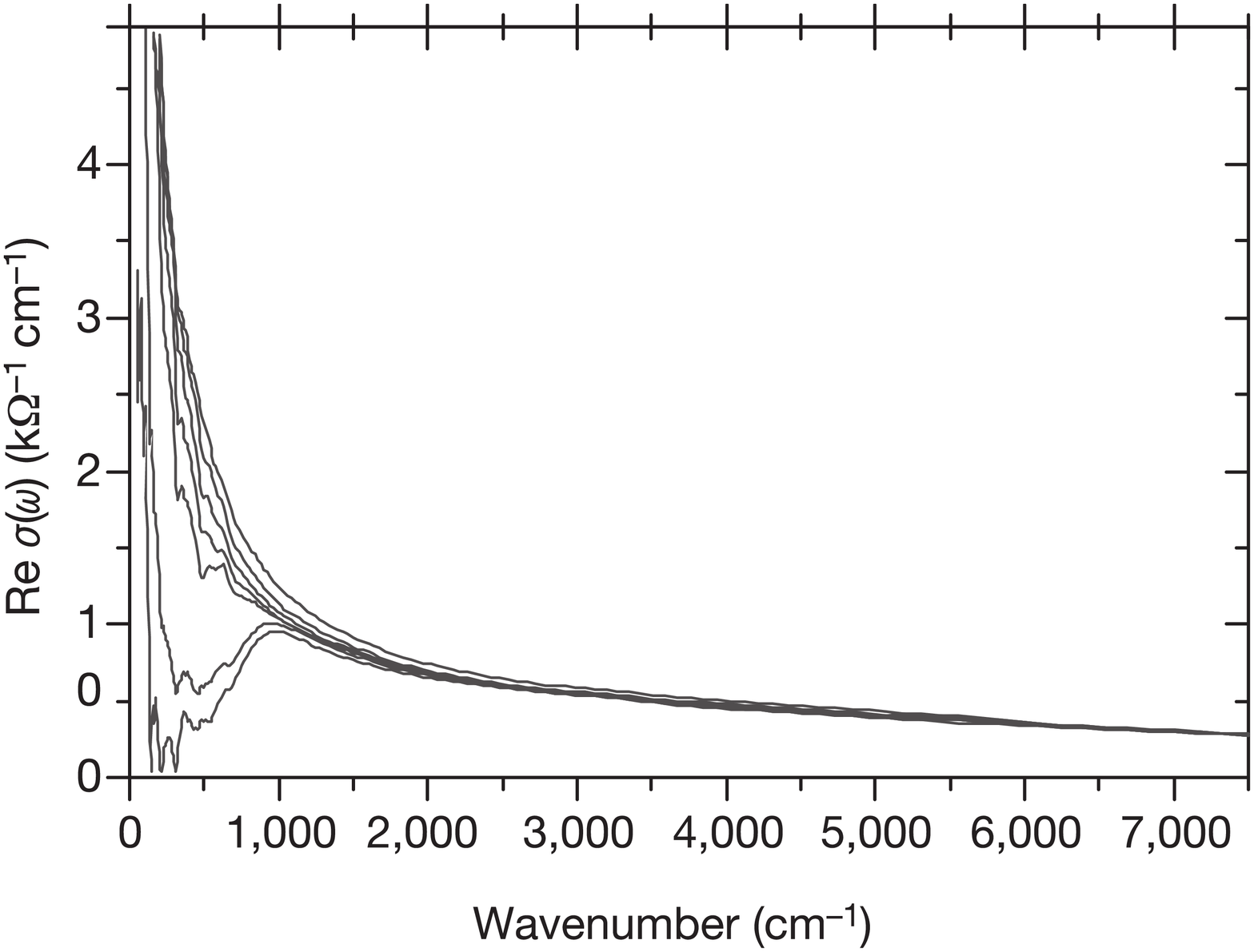} \\
{\scriptsize Fig. 16. The real part of the optical conductivity
along the copper-oxygen planes of
$Bi_2Sr_2Ca_{0.92}Y_{0.08}Cu_2O_{8+\delta}$ for a selected number
of temperatures. Temperatures from left to right are 7, 50, 95,
130, 160, 200, 260 K. The critical temperature is 96 K.
Experimental plot from Ref.~\cite{marel}.}
\end{minipage}
\end{figure}

\end{document}